# Shear thickening in Electrically Stabilized Colloidal Suspensions

Joachim Kaldasch* and Bernhard Senge

Technische Universität Berlin

Fakultät III: Lebensmittelrheologie

Königin-Luise-Strasse 22

14195 Berlin

Germany

Jozua Laven

Eindhoven University of Technology

PO Box 513

5600 MB Eindhoven

The Netherlands

* author to who correspondence should be sent

**Abstract**

A theory is presented for the onset of shear thickening in colloidal suspensions of particles, stabilized by an electrostatic repulsion. Based on an activation model a critical shear stress can be derived for the onset of shear thickening in dense suspensions for a constant potential and a constant charge approach of the spheres. Unlike previous models the total interaction potential is taken into account (sum of attraction and repulsion). The critical shear stress is related to the maximum of the total interaction potential scaled by the free volume per particle. A comparison with experimental investigations shows the applicability of the theory.

**1. Introduction**

A number of theoretical attempts were made in the past to explain shear thickening in stabilized colloidal suspensions. There are two main proposals for the source of shear thickening in these systems: One is that short-range lubrication forces are responsible for the formation of large shear induced density fluctuations, known as hydroclusters (Brady and Bossis (1985), Bender and Wagner (1989), Melrose (2003), Melrose and Ball (2004)). The second is that shear thickening is related to an order-disorder transition, where an ordered, layered structure becomes unstable above a critical shear rate (Hoffman (1974), Boersma et. al. (1990),(1995), Hoffman (1998)). Previous rheological models on shear thickening are based on effective hard spheres, including the mutual repulsion between the particles.

Unlike earlier models we want to present a rheological model on shear thickening that takes into account the van der Waals attraction. For these suspensions the two-particle interaction can be described by the Derjaguin –Landau- Verwey- Overbeek - Theory (Derjaguin and Landau (1941), Verwey and Overbeek (1948)). The standard DLVO-theory involves a particle-particle interaction potential as schematically displayed in Fig.1. It consists of a primary minimum of the potential at small distances, due to the Van der Waals attraction. The formation of an electric double layer around the particles leads to a mutual repulsion, which stabilizes the colloidal suspension. For decreasing surface potentials or increasing temperatures however thermal fluctuations may overcome the electrostatic repulsion leading to the coagulation of the suspension.

The main goal of this paper is to present a theory that determines the critical stress for the onset of shear thickening. After deriving the model, its applicability will be tested by comparing the model predictions with an experimental investigation on aqueous suspensions.

## 2. The Model

We want to consider a concentrated, electrically stabilized, colloidal suspension and suppose that the particle interaction can be described by the superposition of an electrostatic potential $U_{el}(h)$ of the particles and the Van der Waals attraction $U_{vdW}(h)$:

$$U(h) = U_{el}(h) + U_{vdW}(h)$$

(1)

as displayed in Fig.1.

The electrostatic repulsion between two particles is due to the overlap of the electric double layer. In numerous cases, a simple equation derived by Hogg, Healy and Fuerstenau (1966) was found to be a suitable approximation for the calculation of the interaction energy between two charged particles with radii $a_1$ and $a_2$ and relatively small surface potentials $\Psi_1$ and $\Psi_2$

$$U_{HHF}(h) = \frac{\pi \varepsilon_0 \varepsilon_r a_1 a_2}{a_1 + a_2} \left\{ 2\Psi_1 \Psi_2 \ln\left(\frac{1+e^{-\kappa h}}{1-e^{-\kappa h}}\right) \pm (\Psi_1^2 + \Psi_1^2)\ln(1-e^{-2\kappa h}) \right\}$$

(2)

where the parameters $\varepsilon_0$ and $\varepsilon_r$ are the absolute and relative dielectric constants, and $\Psi_0$ is the surface potential. The Debye reciprocal length $\kappa$ is defined by:

$$\kappa = \sqrt{\frac{2C_S N_A z^2 e_0^2}{\varepsilon_0 \varepsilon_r k_B T}}$$

(3)

Here, $k_B$ is the Boltzmann constant, $T$ the temperature, $e_0$ is the elementary electric charge, $N_A$ the Avogadro- number, $z$ the ionic charge number and $C_S$ the molar salt concentration. Eq.(2) was derived under the assumption that the distribution of charge or potential is uniform. The plus sign refers to an approach of the colloidal particles with a constant surface potential. This condition implies that the approach is slow and equilibrium can be established between the ions on the surface and the bulk. The negative sign refers to an approach between the particles, where the surface charge density remains constant. For the case of a monodisperse suspension ($a=a_1=a_2$; $\Psi_0=\Psi_1=\Psi_2$) and a constant potential approach Eq.(2) becomes:

$$U_{CP}(h) = 2\pi \varepsilon_0 \varepsilon_r a \Psi_0^2 \ln(1 + e^{-\kappa h})$$

(4)

and for a constant surface charge approach

$$U_{CC}(h) = -2\pi \varepsilon_0 \varepsilon_r a \Psi_0^2 \ln(1 - e^{-\kappa h})$$

(5)

The Van der Waals attraction between two spheres can be taken into account by

$$U_{vdW}(h) = -\frac{Aa}{12h}$$

(6)

where A is the effective Hamaker constant of the solvent-particle combination.

In equilibrium the particles are distributed more or less randomly with a typical surface to surface distance at equilibrium, $h_0$, which is governed by the volume fraction $\Phi$:

$$h_0(\Phi) \cong 2a\left(\left[\frac{\Phi_m}{\Phi}\right]^{1/3} - 1\right)$$

(7)

where $\Phi_m=0.64$ defines the maximum packing density of a randomly packed hard sphere suspension.

The application of a shear flow to the system induces a deviation of surface to surface distances $h$ between the colloidal particles from the equilibrium value. The maximum compression of the gap between two neighbor particles occurs, when a particle pair is arranged along the compression axis of the sheared system. Within a mean field model, as presented here, the stress $\sigma$ pushing the two particles together is equal to the macroscopic stress of the medium

$$\sigma = \eta(\dot{\gamma})\dot{\gamma}$$

(8)

where $\dot{\gamma}$ is the shear rate and $\eta(\dot{\gamma})$ the shear-rate dependent viscosity of the suspension.

The key idea of this paper is to assume that particles may overcome the mutual electrostatic repulsion. As shown by Ogawa et. al. (1997) the frequency $f$ of density fluctuations in stabilized suspensions is governed by an activation process. According to Eyrings transition state theory [Eyring (1936)], $f$ must have the form of an Arrhenius law:

$$f \sim \exp\left(-\frac{U_B - U_S}{k_B T}\right)$$

(9)

Confining the model to a two-particle consideration the energy barrier formed by the interaction potential is given by:

$$U_B = U(h_{max}) - U(h_0)$$

(10)

where $h_{max}$ is the gap at the maximum of the interaction potential determined by

$$\frac{\partial U(h)}{\partial h} = 0; \frac{\partial^2 U(h)}{\partial h^2} < 0$$

(11)

(Fig.1). For a stabilized suspension in equilibrium we want to assume that $U_B \gg k_B T$.

$U_S$ is a bias potential given by the product of an activation volume $V^*$ and the stress pushing the two particles together

$$U_S = \sigma V^*$$

(12)

$V^*$ represents the volume associated with this activation process, which is of the order of the free volume per particle (Ogawa et. al. (1997)):

$$V^* = \frac{4}{3}\pi a^3 \frac{\Phi_m}{\Phi}$$

(13)

The stress-induced change of the frequency $f$ can be neglected as long as the applied shear stress is small, but increases when the bias potential comes into the order of the maximum of the repulsive potential. We want to associate the critical stress $\sigma_C$ with the maximum of the frequency $f$ at $U_S=U_B$:

$$\sigma_C = \frac{U_B}{\frac{4}{3}\pi a^3 \frac{\Phi_m}{\Phi}}$$

(14)

Note that this critical shear stress determines the initial state of the shear thickening instability. The formation of large shear induced density fluctuations (hydroclusters) can be expected along the compression axis with increasing shear stresses, associated with a considerable increase of the viscosity.

For small shear rates the electrostatic repulsion of two approaching particles can be described by the constant potential approximation. However with increasing shear rates the approaching time of the particles may be comparable with the relaxation time of the electric double layer. Between two approaching spheres the characteristic convective time of the double layer at the transition is given by $\tau_C = (v\kappa)^{-1}$. The approaching velocity $v$ of the particles along the compression axis can be estimated by equating the external mean field force on a particle with the viscous resistance of two approaching particles:

$$\sigma_C \pi a^2 = 6\pi\eta_S a \frac{a}{h} v$$

(15)

where $\eta_S$ is the solvent viscosity. Approximating $h \sim \kappa^{-1}$, we find for the convective time scale at the critical stress:

$$\tau_C \sim \frac{6\eta_S}{\sigma_C}$$

(16)

The diffusive time scale for the lateral relaxation of the double layer due to the squeezing of the gap between two approaching spheres can be estimated:

$$\tau_D \sim \frac{(2a)^2}{D}$$

(17)

Here D is the diffusion constant given by

$$D = \frac{k_B T}{6\pi \eta_S a_i}$$

(18)

where $a_i$ is the radius of the ions. The corresponding Péclet number can be written as the ratio between the convective and the diffusive time scales:

$$Pe = \frac{\tau_C}{\tau_D} = \frac{6\eta_S D}{4\sigma_C a^2}$$

(19)

When *Pe=O(1)* the double layer has not enough time to continuously adapt to a constant surface potential. We can determine a critical stress

$$\sigma_C^* = \frac{6\eta_S D}{4a^2}$$

(20)

which discriminates between a constant potential and a constant charge description of Eq.(14). Thus we finally obtain, that the critical stress is governed by:

$$\sigma_C = \begin{cases} \dfrac{U_{CP}(h_{max}) + U_{vdW}(h_{max}) - U_{CP}(h_0) - U_{vdW}(h_0)}{\dfrac{4}{3}\pi a^3 \dfrac{\Phi_m}{\Phi}} & \sigma_C \ll \sigma_C^* \\ \dfrac{U_{CC}(h_{max}) + U_{vdW}(h_{max}) - U_{CC}(h_0) - U_{vdW}(h_0)}{\dfrac{4}{3}\pi a^3 \dfrac{\Phi_m}{\Phi}} & \sigma_C \gg \sigma_C^* \end{cases}$$

(21)

which is the main result of this paper.

## 4. Comparison with the Experiment

In this section we want to compare the model with experimentally obtained results of electrically stabilized suspensions. Most of the published work on model systems was carried out on hard sphere or effective hard sphere suspensions, where the attraction between the colloidal particles is small. Only a limited number have given a comprehensive characterization of their colloidal particles, which is necessary for the comparison with the presented model.

A study on aqueous alumina suspensions was recently performed by Zhou et. al. (2001). The suspensions were investigated at a constant volume fraction $\Phi=0.56$ and different pH-values, at a salt concentration of 0.01 M $KNO_3$. The samples exhibited a jump-like increase of the viscosity at a critical stress. The characteristic properties of the samples are summarized in Table 1.
Also given in Table 1 is the critical stress of an effective hard sphere model proposed by Maranzano and Wagner (2001), Lee and Wagner (2003):

$$\sigma_C = 0.024 \frac{k_B T \kappa a \Psi^2}{a^2 l_b}$$

(22)

where $l_b$ is the Bjerrum length defined by:

$$l_b = \frac{e^2}{4\pi \varepsilon_0 \varepsilon_r k_B T}$$

(23)

and $\Psi = e\Psi_0/(k_B T)$. Eq.(22) overestimates the critical stresses compared to the experimental data given in Table 1. It demonstrates that this effective hard sphere model is not applicable to the aqueous suspensions.

For the calculation of the attractive contribution between the particles, the Hamaker constant was taken to be $A=4.1*10^{-20}$ $J$ (Lyklema (1991)). The diffusion constant is $D=9.4 \cdot 10^{-10}$ $m^2$ $s^{-1}$ for the large $NO_3$-ions, with an ion radius $a_i=0.23nm$. The experimentally obtained critical stresses and the stresses obtained by Eq.(21) are displayed in Fig.2. The error bars indicate the finite resolution of the measurement. The experimental critical stresses are distributed between the upper bond, the curve calculated under the condition of constant charge critical stress, and the lower bond, the curve for a constant potential critical stress. Also displayed is the stress $\sigma_C^* \approx 8Pa$.

As can be seen from Fig.2 only the sample with the smallest critical shear stress fulfills the condition $\sigma_C < \sigma_C^*$ and is close to the theoretically expected value for the critical shear stress at constant potential. For increasing surface potentials, the experimental results gradually tend towards the constant surface charge critical stress, but saturate close to a curve given by Eq.(14), with $U_B = 2U_{CP}$. This unexpected result does not contradict our model Eq.(21), since it is only valid in the limits of a constant potential and constant charge approximation.

Based on an investigation of the yield stress of the coagulated aqueous suspensions, Zhou et. al. (2001) reported the fact that the average surface-to-surface distance in the coagulated state is between *2.3* and *2.6 nm*, insensitive to the particles size, volume fraction and suspension pH. This result indicates that the particles are rather loosely bounded in the coagulated state. Evaluated in Table 1 is the energy barrier $U_{bound}= U_{CP}(h_{max})-U_{CP}(h)$ of the bounded state for a surface-to-surface distance *h=1nm*. The energy barrier $U_{bound}$ varies between *1 $k_BT$* and *28 $k_BT$*.

For low energy barriers of the bounded state the particles are expected to relax into separated particles by thermal excitations, if the applied shear stress becomes smaller than the critical shear stress. For higher energy barriers the particles are expected to be kept in the bounded state, even below the critical shear stress. Experimentally this hysteresis was found by Zhou et. al. (2001) in the stress response of the samples between increasing and decreasing applied shear stresses.

If $U_{bound}$ is of the order of $k_BT$, the presented theory depicts reversible shear thickening. If $U_{bound} \gg k_BT$, the model describes the viscosity increase associated with a transition into an irreversible coagulated state of the colloidal particles. Note that in the latter case the model describes the initial states of an orthokinetic coagulation in dense suspensions (Smoluchowski (1912), Kruyt (1952), Friedlander (2000)), where disintegration of coagulated structures can be neglected .

## 3. Conclusion

The comparison of an effective hard sphere model of shear thickening, proposed by Maranzano and Wagner (2001), with experimental critical stresses indicates that the van der Waals attraction between the particles cannot be neglected.

The activation model presented here includes the total interaction potential (sum of attraction and repulsion). The relaxation properties of the electric double layer define two critical shear stresses. The critical stresses obtained from an investigation on aqueous suspensions are in agreement with the model predictions. They are located within the upper bond, given by the critical shear stress at constant surface charge, and the lower bond, determined by the critical shear stress at constant surface potential.

We want to emphasize that the present model is not applicable to effective hard sphere suspensions, like sterically stabilized suspensions, since the interaction potential contains no maximum.

Our framework provides an estimation of the critical shear stress of suspensions containing monodisperse colloidal particles with non-neglecting Van der Waals attraction, which is, up to our knowledge, the first instance to do so.

**Tables**

| Suspension | a [μm] | C [mol/m$^3$] | ζ [mV] | pH | T [K] | $\varepsilon_r$ | κa | U$_{Bound}$ [k$_B$T] | σ$_C$ [Pa] Zhou et. al. | σ$_C$ [Pa] Eq.(22) |
|---|---|---|---|---|---|---|---|---|---|---|
| Alumina in water AKP 15L | 0.425 | 10 | 96 | 5 | 293 | 80.37 | 138 | 28 | 50 | 1508 |
| Alumina in water AKP 15L | 0.425 | 10 | 90 | 5.5 | 293 | 80.37 | 138 | 14 | 41 | 1325 |
| Alumina in water AKP 15L | 0.425 | 10 | 80 | 6 | 293 | 80.37 | 138 | 1 | 32 | 1047 |
| Alumina in water AKP 15L | 0.425 | 10 | 72 | 6.4 | 293 | 80.37 | 138 | 1 | 15 | 848 |
| Alumina in water AKP 15L | 0.425 | 10 | 64 | 6.8 | 293 | 80.37 | 138 | 9 | 6 | 670 |

**Table 1.** Characteristic data of the alumina suspensions investigated by Zhou et. al. (2001), where *Φ=0.56, z=1,* the Hamaker constant *A=4.1\*10$^{-20}$ J* and *l$_b$*=6.9 10$^{-10}$ m.

**Figures**

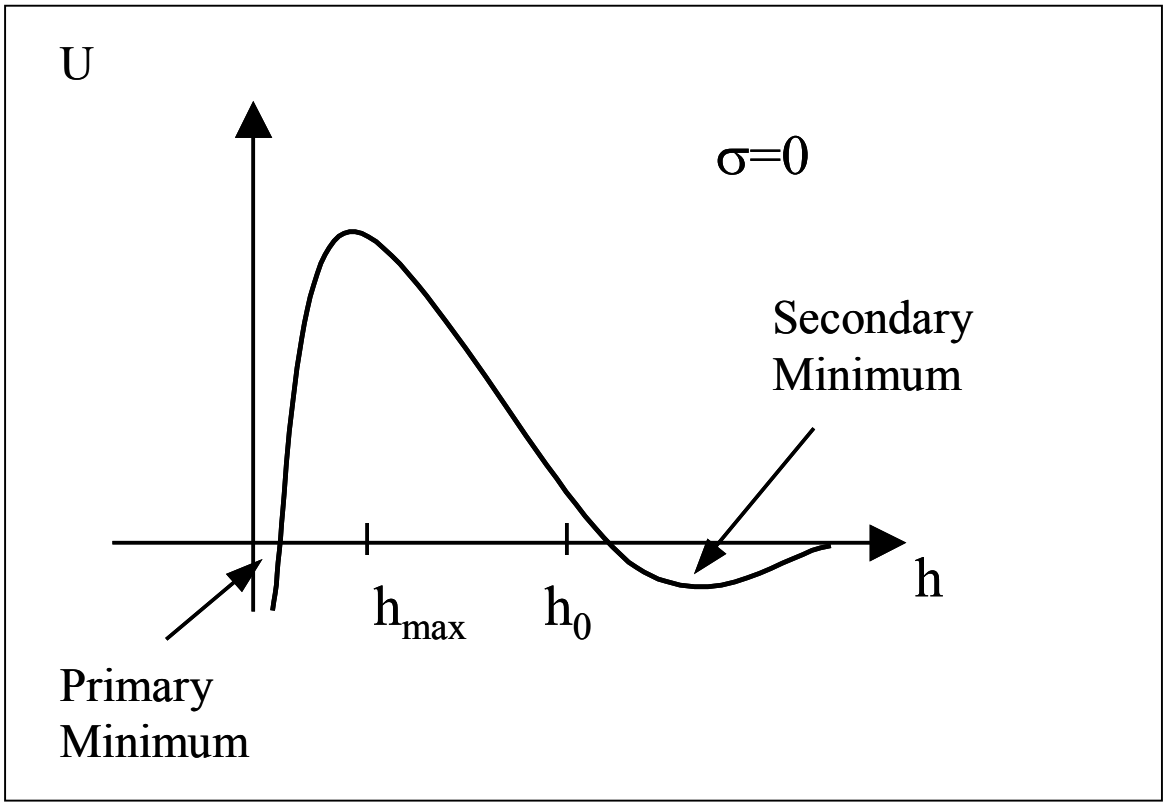

**Figure 1.** Schematic representation of the DLVO interaction potential $U$ as a function of the two-particle surface-to-surface distance $h$.

- 17 -

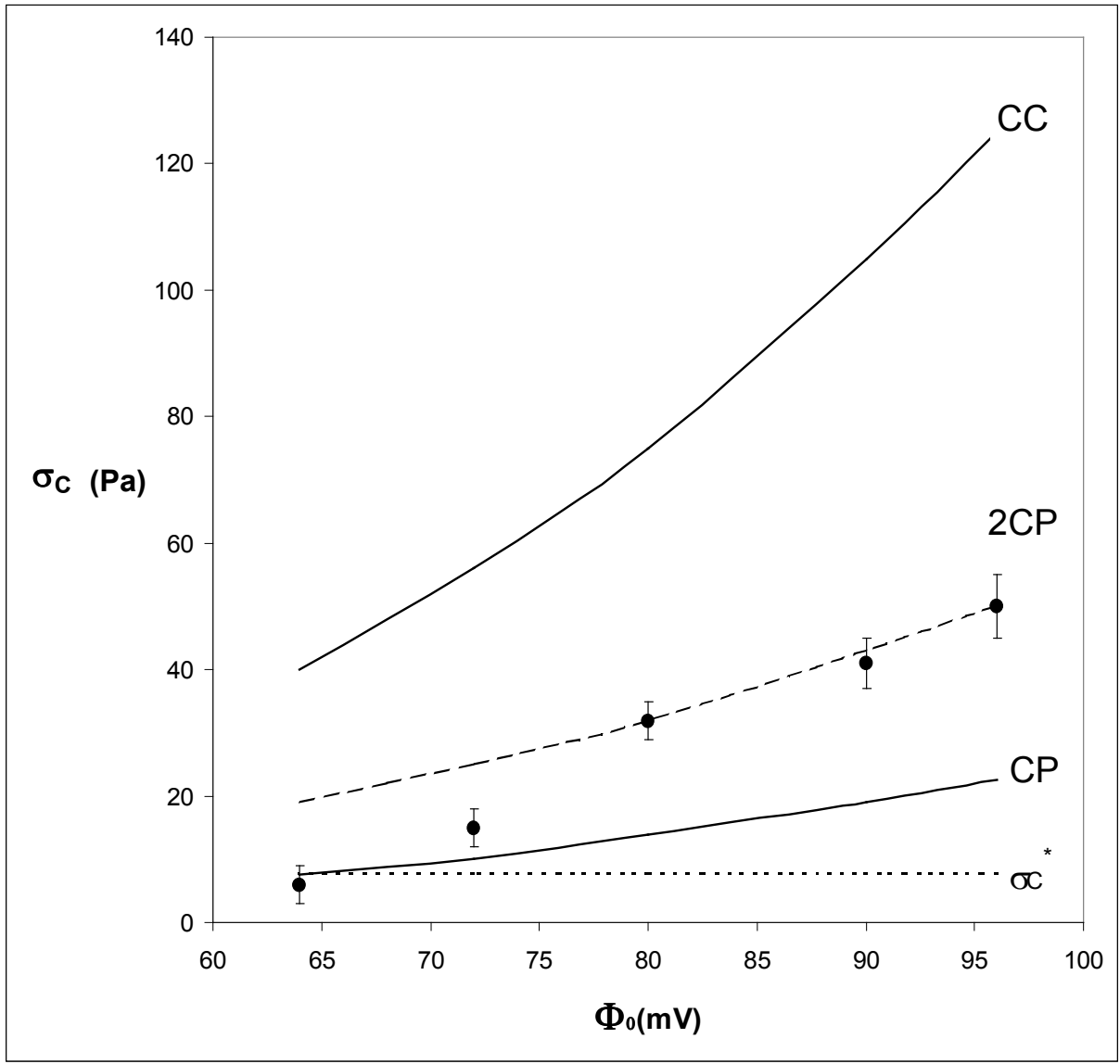

**Figure 2.** Comparison of the model with the data obtained by Zhou et. al. (2001). The points are the critical stresses found for the alumina suspensions. The lines represent: "CC" the constant surface charge critical stress, "CP" the constant surface potential critical stress, "2CP" twice the constant surface potential critical stress (dashed line) and "$\sigma_C^*$" is determined by Eq.(20) (dotted line).